\documentclass[aps,superscriptaddress,prl,reprint]{revtex4-2}

\usepackage{graphicx}
\usepackage{siunitx}

\usepackage{threeparttable}
\usepackage{booktabs}
\usepackage[version=4]{mhchem}
\usepackage{gensymb}

\begin{document}

\title{Flat band in the multiband metal MnSb$_2$}

\author{Carl Jonas Linnemann}

\affiliation{Department of Chemistry, Center for Sustainable Energy Materials, Aarhus University, 8000 Aarhus C, Denmark}

\author{Davide Ceresoli}

\affiliation{Consiglio Nazionale delle Ricerche - Istituto di Scienze e Tecnologie Chimiche “G.
Natta”, via Golgi 19, 20133, Milano, Italy}
\author{Kim-Khuong Huynh}
\email{hkkhuong@chem.au.dk}
\author{Martin Bremholm}
\email{bremholm@chem.au.dk}
\affiliation{Department of Chemistry, Center for Sustainable Energy Materials, Aarhus University, 8000 Aarhus C, Denmark}

\date{\today}

\begin{abstract}

We report the realization of a metallic flat band in the marcasite compound MnSb$_2$.
The material was synthesized under high pressure, and its crystal structure was established by X-ray diffraction.
Electrical and thermoelectric transport measurements reveal signatures of a flat band near the Fermi level that persist across three structural, magnetic, and electronic transitions.
These observations are supported by extensive density-functional-theory calculations.
Related marcasite diantimonides, including FeSb$_2$ and CrSb$_2$, also host flat bands, but these lie far from the Fermi level and therefore do not produce a metallic flat-band state.
MnSb$_2$ therefore emerges as the only known metallic marcasite antimonide in which a flat band lies in the vicinity of the Fermi level.

\end{abstract}
\maketitle

\paragraph{Introduction--}Flat band materials have gained large interest due to their interesting physical properties such as strongly correlated electrons, superconductivity and non-Fermi-liquid behavior. 
They can arise due to destructive interference of orbitals in special lattice types such as the Kagome lattice, due to enlarged unit cells in Moiré systems, or due to strong electron correlations in materials containing $f$- or $d$-electrons~\cite{Flat_band_review,Flat_band_ferm_liq}.

The last category of flat band physics is particularly important for the family of transition metal diantimonides, TMSb$_2$, where TM is a $3d$-metal among Cr, Mn, Fe and Ni~\cite{marcasite_structures, MnSb2, Arsenopyrites_high_T}.
In these compounds, the flat bands originate from the orientation of the $3d$-orbitals in the marcasite structure.
The most famous members of this family are FeSb$_2$ and its doped phases, which are known for showing colossal thermopower and as candidates for Kondo insulators  and altermagnets \cite{PetrovicFeSb2,FeSb2_colossal_seebeck,FeSb2_MAZIN,FeSb2_Sn_doping,FeSb2_Co_doping, CrFeSb2_doping_study}.

\begin{figure}
    \includegraphics[]{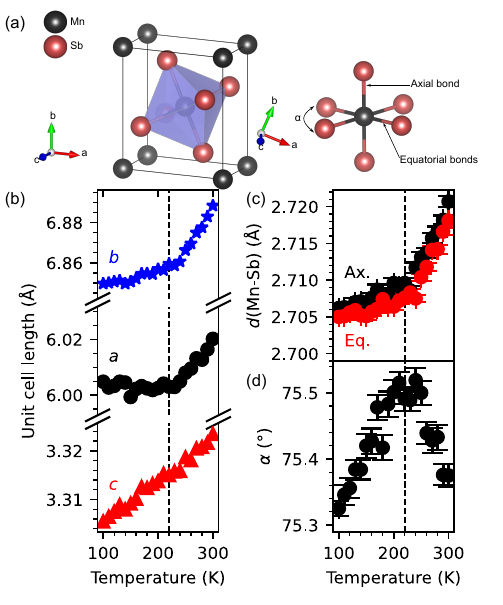}
    \caption{\label{fig:MnSb2_struct} (a) Crystal structure of MnSb$_2$, showing the MnSb$_6$ octahedron with axial and equatorial Mn-Sb bonds. 
    (b) Extracted unit cell parameters from SCXRD showing a kink at 220 K. (c) Mn-Sb bond lengths (d) Bond angle $\alpha$.}
\end{figure}

The marcasite structure, which has the space group $Pnnm$ (Fig. \ref{fig:MnSb2_struct} (a)), allows a configuration in which the TM's $3d_{zx}$ and $3d_{zy}$ orbitals point towards the faces of the TMSb$_6$ octahedra and therefore become non-bonding~\cite{BjoernFeSb2}.
These rather localized $d$ orbitals result in a flat band observed at about  $\SI{0.1}{\electronvolt}$ below $E_\mathrm{F}$ in FeSb$_2$, as verified by angle-resolved photoemission spectroscopy (ARPES) \cite{FeSb2_ARPES}.
Although the flat band is realized, its large distance from $E_\mathrm{F}$ hinders a direct observation of its behavior in transport properties.
The same challenge also appears in CrSb$_2$, which on the other hand has the flat band above the Fermi level~\cite{CrSb2_magnon}. MnSb$_2$, which has remained a system hidden from experimental studies, is a fortunate case where the flat band may be realized at the Fermi level due to the number of valence electrons.

The coexistence of colossal thermopower, intriguing magnetic behavior, and the flat band has motivated extensive studies of TMSb$_2$ compounds.
However, the lack of knowledge for MnSb$_2$ has hindered a thorough understanding of TMSb$_2$. 
This compound can only be synthesized using a high pressure approach.
It was first synthesized by Takizawa et al., who noted that the magnetic susceptibility was temperature independent, but no further physical properties were investigated~\cite{MnSb2}. 
More recently, in a study on possible altermagnetism in doped FeSb$_2$, theoretical calculations on MnSb$_2$ were also published, showing that the AFMe order, as also favored in CrSb$_2$, is more stable than the altermagnetic state~\cite{FeSb2_MAZIN}.

Here, we investigate the temperature dependent structural and electronic properties of MnSb$_2$ using single-crystal X-ray diffraction (SC-XRD), electrical transport measurements performed on a polycrystalline sample,  and density functional theory (DFT) calculations. Our results indicate the presence of a flat band in the vicinity of the Fermi level, establishing MnSb$_2$ as the first metallic marcasite compound with a flat band located close to the Fermi level.
The measurements of the Seebeck coefficient, resistivity, and heat capacity reveal two transitions at 110 K and 220 K, whereas an additional electronic transition appears at around 30 K, at which the Nernst coefficient switches its sign. These findings position MnSb$_2$ as a unique member of the TMSb$_2$ family, fill the gap in the series, and strengthen marcasites as a platform for exploration of flat‑band-driven physics.

\paragraph{Synthesis and crystal structure --}
We employed a high-pressure ($\SI{4}{\giga\pascal}$) and high-temperature ($\SI{800}{\celsius}$)~\cite{COMPRES} synthesis to obtain the marcasite phase of \ce{MnSb2} \footnote{See Supplemental Material at https://doi.org/10.1103/cp97-khvn}.
Elaborated structural investigations using both single-crystal (SC-) and powder X-ray diffraction (P-XRD) analyses show that the product of the high-pressure synthesis is of high crystallinity and \SI{98}{\percent} phase pure (Fig.~S1)~\cite{Note1, cryspro,Olex2,Shelxl,Shelxt,Fullprof,vesta,Matfraia,plaid}.
The only detectable impurities are \ce{Sb}, i.e. $\approx$\SI{2}{wt \percent} and MnSb which is barely visible in the XRD but was observed through magnetic measurements~\cite{MnSb_satmag,Note1}. Our X-ray map (Fig.~S1~\cite{Note1}) shoved that the MnSb and Sb impurities are randomly distributed through the sample and no channels of MnSb or Sb could be observed, even when projected into a 2D map~\cite{Note1, pippard,Sb_MR}.

The structure solution of MnSb$_2$ obtained from SCXRD refinement (Fig.~\ref{fig:MnSb2_struct}) replicates the marcasite structure and, at the same time, verifies the first structural model proposed by Takizawa et al. through Rietveld refinement pf PXRD data~\cite{MnSb2}.
In the marcasite-type structure the Mn atoms form a body-centered sublattice with the Sb-atoms forming dimers in the $ab$-plane. Mn is coordinated in a distorted octahedra to 6 Sb atoms with two distinct bonds.
There are two axial bonds and four equatorial bonds and the angle between the equatorial bonds is named $\alpha$.
Several studies have shown that these bond lengths and angles are highly dependent on the $d$-electron configuration.
In the case of FeSb$_2$, the equatorial bonds are longer than the axial bonds, whereas the opposite is observed for CrSb$_2$ and MnSb$_2$, indicating that their electronic configuration might be similar~\cite{BjoernFeSb2,MnSb2,<CrSb2_unitcell>,GOODENOUGH_marcasites}.
Temperature resolved SCXRD measurements (Fig.~\ref{fig:MnSb2_struct}(b-d)) indicate an anomaly at $T^\ast \approx \SI{220}{\kelvin}$, below which the $a-$ and $b-$ axis stiffen.
At $T < T^\ast$, the $a$- and $b$-lattice parameters are virtually constant with decreasing temperature, being sharply contrast to the linear variation of ${c}$.
Accordingly, at $T^\ast$, both equatorial and axial Mn-Sb bond lengths exhibit a kink, and the equatorial Sb-Mn-Sb bond angle, $\alpha$, displays a maximum.
No change of the Sb-Sb bond lengths was observed within the uncertainty of the measurement.

\paragraph{Electronic structure --}
Using the experimentally determined crystal structure, we proceeded with the calculation of the electronic structure~\cite{QE1,QE2, GBRV,PBEsol,PBEsol, PAOFLOW2,Fermi_surfer,Note1}.
Energetic optimizations carried out across various candidates for the ordering of Mn's spins suggest an antiferromagnetic ground state (Table S2~\cite{Note1}), named AFMe as illustrated in Fig. 1a of Ref.~\cite{Kuhn2013}, together with three other possible antiferromagnetic orders. The AFMe order is the lowest in energy and is consistent with earlier calculations~\cite{FeSb2_MAZIN} with a magnetic moment of 2.5 $\mu_\mathrm{B}$ for Mn. It can be visualized as a 1x2x2 supercell of the marcasite structure where the moments are coupled ferromagnetically within the (011) plane of the magnetic cell while adjacent planes are coupled antiferromagnetically~\cite{<CrSb2_neutron>}.

\begin{figure}[hb]
    \includegraphics[]{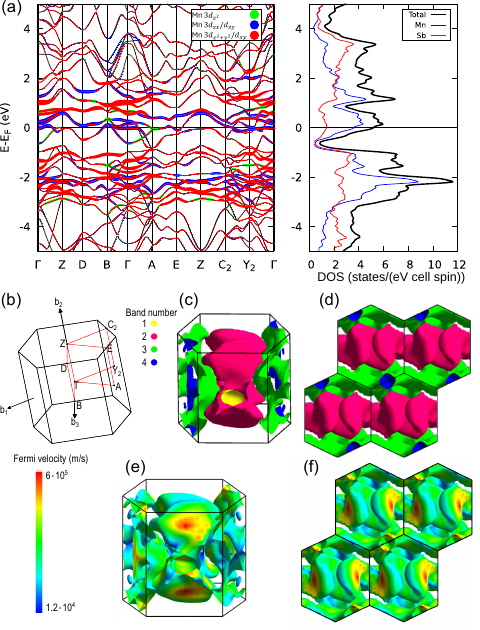}
    \caption{\label{fig:MnSb2 bands} (a) Calculated electronic band structure for the magnetic AFMe superstructure of MnSb$_2$, with colors indicating the contributions from Mn $3d$ orbitals, and the total and atomic projected density of states. 
    (b) Brillouin zone and the path used in the band structure (c-d) Complete Fermi surface, with colors marking the contributions from the different bands. (e-f) Fermi surface colored according to the Fermi velocity, where especially band 2 is seen to possess both slow and fast parts.}
\end{figure}

The band structure of the energetically selected AFM ground state (Fig.~\ref{fig:MnSb2 bands}(a)) displays a remarkable flat band positioned in the vicinity of the Fermi level ($E_{\mathrm{F}}$).
Originating mainly from the Mn $3d_{zx}$ and $3d_{zy}$ orbitals, the flat band extends over a wide range in the momentum ($k$-) space, being of decisive importance for the physical properties of \ce{MnSb2}.
When comparing with the band structures of the other magnetic orderings (Fig.~S3~\cite{Note1}), it is seen that the flat band is present even in the non-magnetic band structure, indicating that the flat band is an intrinsic feature of \ce{MnSb2}. Furthermore, the flat bands are present in the two most energetically stable AFM orders, AFMe and AFMo, where it for AFMo is around 0.2 eV below the Fermi level. Xu et al. showed that the magnetic structure of MnSb$_2$ is incommensurate~\cite{MnSb2_new}, and as such making a definite conclusion about the exact position of the flat bands from the DFT calculations is not possible and would require ARPES studies on single crystals of MnSb$_2$.
The flat bands of \ce{MnSb2} bear great similarities with those of \ce{FeSb2}, the latter were experimentally observed by both ARPES and chemical bonding analyses~\cite{FeSb2_ARPES,BjoernFeSb2}.
The other bands at or close to the Fermi energy have mainly contributions from the $3d_{x^2+y^2}/d_{xy}$ orbitals, while a smaller contribution from the $3d_{z^2}$ orbitals is also present.

The DOS is similar to the one calculated by Stone et al. for \ce{CrSb2}~\cite{CrSb2_magnon}. For MnSb$_2$, the $E_\mathrm{F}$ is not in the band gap though, but lying higher attributed to the fact that Mn has one more electron than Cr, whereby MnSb$_2$ can be seen as electron doped CrSb$_2$. 
When comparing the band-structures of nonmagnetic CrSb$_2$, MnSb$_2$ and FeSb$_2$ (Fig.~S4~\cite{Note1}), it is seen that the flat band is above $E_\mathrm{F}$ for CrSb$_2$, at $E_\mathrm{F}$ for MnSb$_2$ and below $E_\mathrm{F}$ for FeSb$_2$.
The position of the flat band in MnSb$_2$ can be understood from the fact that the electron count of Mn is between those of Fe and Cr, leading to the Fermi Energy being in the vicinity of the flat band.

The Fermi surface of MnSb$_2$ is shown in Fig~\ref{fig:MnSb2 bands}. 
Four bands cross the Fermi energy, giving rise to two tubular shapes (red and green) as well as two smaller complex open shapes crossing the BZ boundaries (yellow and blue). 
We number the bands as follows: band 1 (yellow), band 2 (pink), band 3 (green) and band 4 (blue). Band 1 is a small hole pocket originating from around B with contributions mainly from the $3d_{zx}/3d_{zy}$ orbitals.
Band 2 and 3 have more complex behavior, and are the bands that travel along the $b_1$ direction, while band 4 mainly travels along the $b_2$ direction. Especially band 2 has many characteristics, and possesses both slow and fast parts as seen in Fig.~\ref{fig:MnSb2 bands} (e-f) where the Fermi surface is colored according to its Fermi velocity.

\begin{figure*}
    \includegraphics[width = \textwidth]{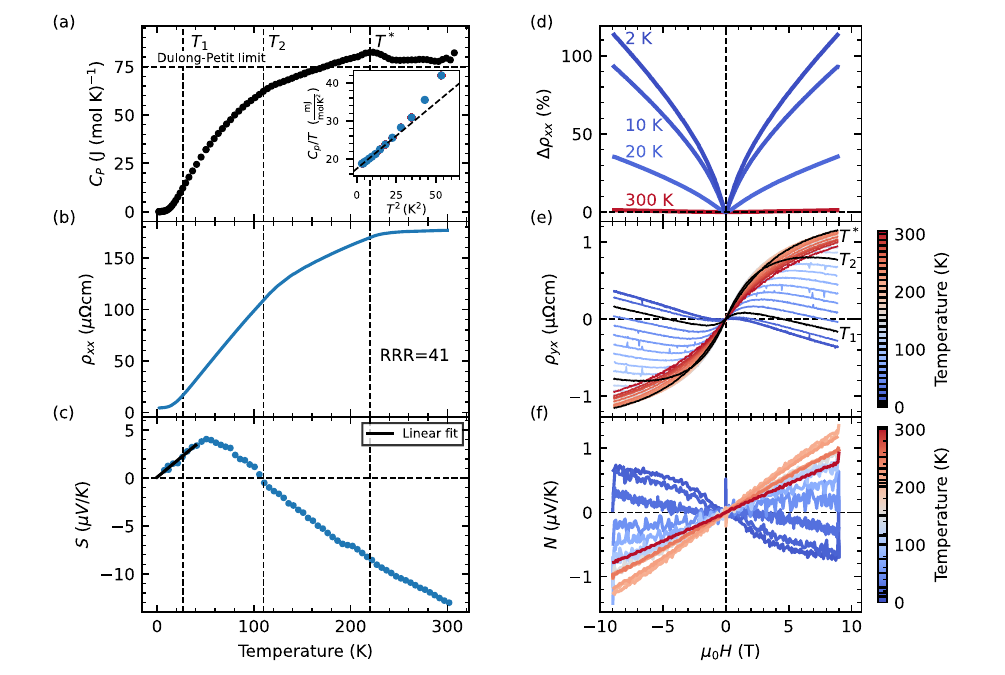}
    \caption{Physical properties of \ce{MnSb2}: (a) Heat capacity. The insert showing the fit to the low temperature heat capacity. (b) Resistivity, $\rho_{xx}$. (c) Seebeck coefficient with a linear fit to the low temperature Seebeck coefficient. (d) Magnetoresistivity at selected temperatures. (e) All measured $\rho_{yx}$. The curves collected at temperatures closest to the observed transitions are shown in black. f) Nernst signal. The vertical lines in (a-c) mark the observed transitions, while in (e) the lines serve as a guide to the eye.}
    \label{fig:Physprop}
\end{figure*}

\paragraph{Physical properties and phase diagram --}
The physical properties of \ce{MnSb2} show that it is a non-trivial metal showing multiple transitions (Fig.~\ref{fig:Physprop}).
The structural anomaly at $T^\ast$ is also clearly seen in the measured heat capacity, $C_P$, and resistivity, $\rho_{xx}$.
An additional anomaly also appears at $T_2 \approx \SI{110}{\kelvin}$ where the Seebeck coefficient, $S$, switches its sign and the heat capacity and resistivity possess a kink while $T_1\approx\SI{30}{\kelvin}$ is due to the Nernst signal, $N$, switching its sign and is not visible in the measured heat capacity.

For $T > T^{\ast}$, the resistivity (Fig.~\ref{fig:Physprop} (a)) shows a weak $T$-dependence, the magnetoresistance is negligible, and the Nernst signal is small and linear with magnetic field.
Interestingly, in this region, the positive sign of the Hall resistivity, $\rho_{yx}$, is sharply in contrast to the negative sign of the Seebeck coefficient.
Whereas $\rho_{yx}$ probes the cyclotron motions of electrical carriers in response to magnetic field and thus is dominated by the faster carrier type, the sign of $S$ is decided by that of the major carriers.
We thus interpret the sign difference as originated from the coexistence of a fast, holelike carrier type of minor density and a major electronlike carrier type with a much smaller mobility.
The bend of $\rho_{yx}$ towards a negative slope at the high field regime supports this interpretation.

For $T < T^{\ast}$ the high field $\rho_{yx}$ decreases and bends increasingly more downwards indicating larger contribution from electrons while $N$ begins to flatten while approaching $T_1$.
Below $T_1$ the Nernst signal switches sign, the magnetoresistivity rises sharply and reaches $110\%$ and the Seebeck coefficient decreases linearly towards 0 as seen in several heavy fermion compounds and metals~\cite{Corr_Seebeck}.
The hall effect only has a small linear slope at low fields with a large electron like tail, indicating the presence of fast holes with low carrier concentration as well as slow electrons with a high carrier concentration.
As expected, the heat capacity, $C_P$, flattens out and from fitting $C_P/T=\gamma +AT^2$, where $\gamma$ is related to the DOS at the Fermi energy~\cite{Ashcroft_book,Kittel}, $\gamma$=\SI{17.4(1)}{m\joule/mol\,\kelvin^2} was obtained.
As both Sb and MnSb have a low $\gamma$~\cite{Sb_sommerfeld,MnSb_sommerfeld}, their contribution to the measured $\gamma$ is negligible~\cite{Note1} providing further evidence, that MnSb$_2$ is metallic and possesses a flat band in the vicinity of the Fermi level.
The heat capacity around $T^\ast$ was also measured under magnetic fields, but no obvious changes to the transition were observed (Fig.~S5~\cite{Note1}).

\begin{figure}
    \includegraphics[]{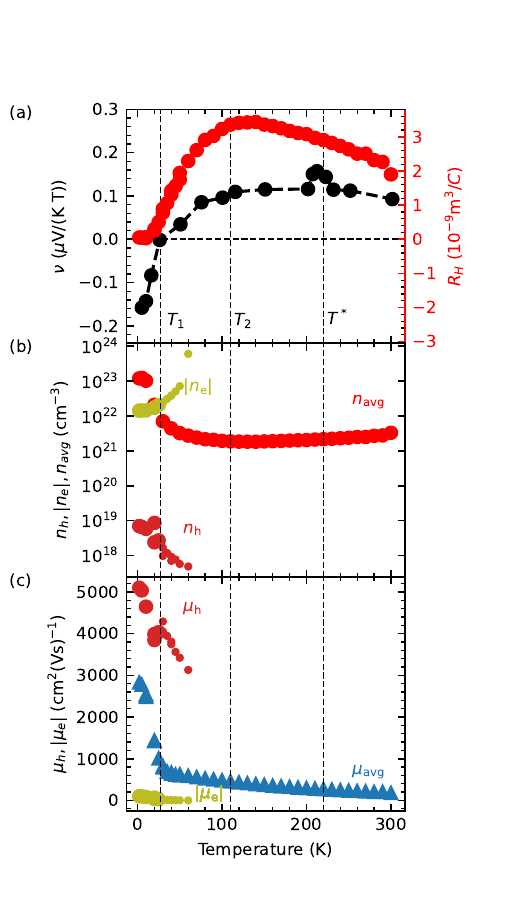}
    \caption{Derived properties from $\rho_{yx}$ and the measurements of the Nernst effect. (a) Nernst coefficient $\nu$ showing a sign switching below 30 K together with the Hall coefficient extracted from a linear fit between $\pm\SI{1}{\tesla}$ to the low field $\rho_{yx}$. (b) Carrier concentrations extracted from $\rho_{yx}$. (c) Mobilities extracted from $\rho_{yx}$ as well as the average mobility extracted from the low field $\rho_{xx}$. The smaller font size above $T_1$ in (b-c) mark the reduced fit quality from the two-band fits.}
    \label{fig:Physprop_derived_2}
\end{figure}

The parameters extracted from the measured transport properties are shown in Fig.~\ref{fig:Physprop_derived_2}. 
The Nernst coefficient, $\nu$,  was derived from a linear fit to the low field value, and the results are seen in Fig.~\ref{fig:Physprop_derived_2}~(a). Above $T_2$ the value is constant and has a similar value to NbSe$_2$~\cite{NbSe2_nernst}, while it decreases below $T_2$ and switches sign at $T_1$ around 30 K. A peak is seen around $T^*$, in accordance with the other observed phase transition.
For the Hall coefficient, $R_H$, extracted by performing a linear fit to the low field $\rho_{yx}$, a maximum around 110 K is seen, where the Seebeck coefficient switches sign, and $R_H$ flattens out to 0 around the sign switch of $\nu$.
Above \SI{30}{\kelvin} the fits with the 2-band model to $\rho_{yx}$ became increasingly unphysical  (Fig.~S6~\cite{Note1}), and therefore only results from fits at low temperatures were included.
A fast hole-like carrier type with low carrier number was seen, as well as a slow, electron like carrier, with high carrier number (Fig.~\ref{fig:Physprop_derived_2} (b-c)). At higher temperatures the carrier numbers and mobilities were estimated from the low field Hall coefficient, and magnetoresistance respectively. The mobility decreases slightly with increasing temperature, while the carrier number is constant, however below 30 K, where $\nu$ switches sign $\mu_\mathrm{avg}$ increases rapidly together with $\mu_\mathrm{h}$.

\paragraph{Discussion--}
From the band structure (Fig.~\ref{fig:MnSb2 bands}(a)) a flat band was observed in the vicinity of $E_\mathrm{F}$, which is in accordance with the observed $\gamma$.
From the electron carrier concentration at 2 K ($n_e=1.4\cdot10^{22}$ \si{cm^{-3}}) and the unit cell volume extracted experimentally at 100 K, the effective mass of the electrons is estimated to be $11m_e$. 
This is similar to what is found for hole doped FeSb$_{2-x}$Sn$_x$, where hole doping reduced the Fermi energy towards the flat band seen in FeSb$_2$~\cite{FeSb2_Sn_doping}.
Thus, both the heat capacity and $\rho_{yx}$ show the presence of heavy (slow) electrons, indicating the presence of a flat band close to the Fermi level.
This is furthermore in accordance with the calculated Fermi surface which both contains small hole pockets as well as larger complex shapes.

The fact that MnSb$_2$ is metallic is peculiar, since both its neighboring compounds FeSb$_2$ and CrSb$_2$ are semiconducting~\cite{Sales_CrSb2,FeSb2_hulliger_Helvetica}.
However, as the DOS of both CrSb$_2$ and MnSb$_2$ are similar and their lattice constants are almost equal~\cite{CrSb2_magnon}, it can be understood by the fact that Mn has one more electron than Cr which thus goes into the conduction band.
Thereby, MnSb$_2$ can be seen as electron doped CrSb$_2$.
Another way of electron doping CrSb$_2$ is by introducing Fe, which has been performed by Hu et al. who studied the electronic phase diagram of \ce{Fe}$_{1-x}$\ce{Cr}$_x$\ce{Sb2}~\cite{CrFeSb2_doping_study}.
For $x=0.5$, which corresponds to the same electron count as \ce{MnSb2}, AFM ordering was observed, although the system was still semiconducting from the resistivity measurements. The Sommerfeld constant for \ce{Fe}$_{0.5}$\ce{Cr}$_{0.5}$\ce{Sb2}, $\gamma=$\SI{10.7}{mJ/mol\,\kelvin^2} is similar to the value for \ce{MnSb2}~\cite{CrFeSb2_doping_study}. 

For the phase transition around 220 K, both the Seebeck and Nernst signal show a kink, the bond angle $\alpha$ has a maximum and the high field $\rho_{yx}$ has a maximum, indicating a phase transition affecting the conducting electrons.
Xu et al. showed that the transition is due to ordering to an incommensurate AFM ground state, where most of their proposed models had little orientation of the magnetic moments along the $c$-axis~\cite{MnSb2_new}. 
The fact that the kink in the lattice parameters is not visible in the $c$-axis but only the other axes, shows that the transition is directional in the $ab$ plane, and it would thus be interesting to study the properties of MnSb$_2$ along different axes if sufficiently large single crystals could be obtained. 
The measured magnetic susceptibility, $\chi$, (Fig.~S5~\cite{Note1}) showed a broad hump above the transition at $T^*$, which is similar to what is observed for CrSb$_2$, where the AFM transition is seen through a small kink around $T_\mathrm{N}$ \cite{Sales_CrSb2}.

At $T_2$ the Seebeck coefficient switches sign which can have several origins, such as a simple switching of carrier type. 
However, from the $\rho_{yx}$ two carrier types were observed and the reason is thus not simple. 
In general, the Seebeck coefficient depends on the number of states above and below the Fermi level, but not at the Fermi level. In contrast, the electrical transport depends on the states at the Fermi level, which is also shown through the Mott formula~\cite{Behnia_book}:
\begin{align}
    S&=-\frac{\pi^2}{3}\frac{k_\mathrm{B}T}{e}\left.\frac{\partial \ln{\sigma}}{\partial E}\right|_{E=E_F}
\end{align}
The electrical conductivity, $\sigma$, depends on both the DOS at the Fermi level as well as the relaxation time, and as such in the constant relaxation time approximation, a slight movement of the Fermi level, especially if a flat band is present in the vicinity of the Fermi level, can change the sign of the Seebeck coefficient, which is hypothesized to occur here.
Furthermore, this can explain the discrepancy between the hall effect and the Seebeck coefficient.
The presence of the finite Nernst signal furthermore verifies the multiband nature of MnSb$_2$ as due to the Sondheimer cancellation, the Nernst signal usually only occurs in materials with more than one charge carrier. 
The Nernst signal can for a two-band metal be shown to be~\cite{Wang_formula, NbSe2_nernst}:
\begin{align}
    N&=-\left.\frac{\pi^2}{3}\frac{k_B^2T}{e}\frac{\partial\tan(\theta_\mathrm{H})}{\partial E}\right|_{E=E_F}\\
    &=S(\tan(\theta_\alpha)-\tan(\theta_\mathrm{H}))\\
    &=S\left(\frac{\alpha_{xy}^h+\alpha_{xy}^e}{\alpha_{xx}^h+\alpha_{xx}^e}-\frac{\sigma_{xy}^h+\sigma_{xy}^e}{\sigma_{xx}^h+\sigma_{xx}^e}\right)
\end{align}
Where $\tan(\theta_\mathrm{H})=\frac{\sigma_{xy}^h+\sigma_{xy}^e}{\sigma_{xx}^h+\sigma_{xx}^e}$ is the Hall angle, $\tan(\theta_\mathrm{\alpha})$ is defined as $\tan(\theta_\mathrm{\alpha})=\frac{\alpha_{xy}^h+\alpha_{xy}^e}{\alpha_{xx}^h+\alpha_{xx}^e}$ where $\alpha_{ij}$ are the elements of the Peltier conductivity tensor for the holes, $h$, or electrons, $e$.
For a one-band metal $\tan(\theta_\alpha)-\tan(\theta_H)=0$ leading to a vanishing Nernst signal.
From this it can be seen that the sign change could be due to a sign change of $\tan (\theta_\mathrm{H})$. However, from our measured Hall angle (Fig.~S6~\cite{Note1}), the Hall angle is smaller than $N/S$ by a factor of 100, and thus the sign switching is due to $\tan(\theta_\alpha)$ switching its sign.
Furthermore, while $\alpha_{xx}^h$ and $\alpha_{xx}^e$ have opposite signs, which would cancel out with the sign of $S$, as $S=\alpha_{xx}/\sigma_{xx}$, $\alpha_{xy}^e$ and $\alpha_{xy}^h$ have the same sign, such that the Nernst signal usually does not switch its sign~\cite{NbSe2_nernst}.
Nevertheless, a sign change for the Nernst signal has been observed in several compounds such as FeSi and Ce$_{1-x}$La$_x$Cu$_2$Si$_2$, and can have different origins such as asymmetric Kondo scattering shown for Ce$_{1-x}$La$_x$Cu$_2$Si$_2$~\cite{Heavy_ferm_Nernst,FeSi_Nernst}. 
As such for MnSb$_2$ it is assumed that an uncommon energy dependence of $\tan (\theta_\mathrm{H})$ leads to a sign switch of $\frac{\partial\tan(\theta_\mathrm{H})}{\partial E}|_{E=E_F}$ in eq.~2 giving $N<0$.
Above $T_1$, where $\nu$ switches sign, the fits to $\rho_{yx}$ with the two band model get unreliable furthermore indicating an electronic phase transition (Fits to the data are shown in Fig.~S6~\cite{Note1}).
A peak appears in the hole mobility, however the absolute values above $T_1$ should not be trusted too much due to the decreasing fit quality.
Furthermore, below 30 K, the concentration of electrons decreases, while the concentration of holes increases, accompanied by a large increase in their mobility. 
This is furthermore verified by the sharp increase in magnetoresistance and thereby $\mu_\mathrm{avg}$ below 30 K.
This indicates that below the electronic transition at $T_1$ the scattering time of the electrons is largely reduced, which can furthermore hint towards an uncommon energy dependence of $\tan (\theta_\mathrm{H})$ and the sign change of $\nu$ is thus attributed to the condensation of MnSb$_2$ into a ground state with an uncommon energy dependence of $\tan (\theta_\mathrm{H})$.

\paragraph{Conclusion--}
The physical properties of MnSb$_2$ were studied showing the presence of two phase transitions manifested in  the heat capacity and other measured physical properties as well as an electronic transition around 30 K.
Both experiments and DFT calculations suggest the presence of a flat band in the vicinity of the Fermi level.
This is in contrast to its neighbouring compounds FeSb$_2$ and CrSb$_2$ which both are semiconducting and possess flat bands below or above the Fermi level respectively.
MnSb$_2$ was shown to be a multi-band metal with a flat band in the vicinity of the Fermi level, leading to the presence of heavy electrons.
Furthermore, hole-carriers with high mobility were observed, leading to a high magnetoresistance at low temperature.
During cooling MnSb$_2$ undergoes several transitions which were manifested in the measured physical properties. Above 220 K it is paramagnetic, below 110 K the Seebeck coefficient switches sign, and below 30 K the Nernst coefficient switches sign and the mobility of the hole carriers rapidly increases.
These results show that MnSb$_2$ is a non-trivial metal with an interesting ground state, motivating further studies on MnSb$_2$ such as attempting to grow millimeter sized single crystals or attempting to stabilize the phase as a thin film to measure its anisotropic properties.
This would also allow for angle-resolved photoemission spectroscopy measurements which could help to illucidate the band structure of MnSb$_2$ in the vicinity of the Fermi level.

\paragraph{Note added in proof --}
During the finalization of the manuscript, we became aware of another study on MnSb$_2$~\cite{MnSb2_new}, which focuses on the magnetic structure of the material.

\begin{acknowledgments}
\paragraph{Acknowledgments--}
We would like to thank Bo Brummerstedt Iversen for his support and discussion, and Kasper Borup and Innokenty Kantor for assistance with the experiments. 
We acknowledge Yong P. Chen for lending us his electronics equipments.
The Independent Research Fund Denmark is acknowledged for funding the project (DFF-FNU, grant no. 1026-00409B) as well as the Villum Foundation (25861), and we further acknowledge the Danish Agency for Science, Technology and Innovation for funding the instrument center DanScatt.
We acknowledge the MAX IV Laboratory for beamtime on the DanMAX beamline under proposal 20241611. Research conducted at MAX IV, a Swedish national user facility, is supported by Vetenskapsrådet (Swedish Research Council, VR) under contract 2018-07152, Vinnova (Swedish Governmental Agency for Innovation Systems) under contract 2018-04969 and Formas under contract 2019-02496. DanMAX is funded by the NUFI grant no. 4059-00009B.
We acknowledge the Carlsberg Foundation (CF20–0364) and the Aarhus University Materials Research Cluster for the Tescan Clara SEM instrument.
\end{acknowledgments}

\bibliography{references.bib}

\end{document}